\documentclass[11pt,a4paper]{article}

\usepackage{amsmath}
\usepackage{amsfonts}
\usepackage{amssymb}
\usepackage[scr=boondoxupr]{mathalpha}
\usepackage{lmodern}
\usepackage{graphicx}
\usepackage[left=2.5cm,right=2.5cm,top=2.5cm,bottom=2.5cm]{geometry}
\usepackage{tensor}
\usepackage{braket}
\usepackage[nottoc]{tocbibind} 
\usepackage{bbm} 
\usepackage{slashed} 
\usepackage{color}
\usepackage{cite}

\usepackage{color} 
\usepackage{hyperref} 
\hypersetup{
    colorlinks=true, 
    linkcolor= black, 
    urlcolor= black, 
    citecolor = black,
    linktoc=all 
}

\allowdisplaybreaks  

\numberwithin{equation}{section} 

\DeclareMathOperator{\tr}{tr}

\newcommand{\dd}[1]{d#1\ }

\newcommand{\WW}{ \mathfrak{F} }

\newcommand{\1}{ { \mathbbm{1} } }

\newcommand{\cd}{ \nabla }

\newcommand{\intl }{\int\limits  }

\begin{document}

\begin{center}
\vspace*{2cm}

{\Large\bf Conformal anomalies in 6d 4-derivative theories: \\ \vspace{0.2cm}
a heat-kernel analysis
}

\vspace{1cm}

{Lorenzo Casarin
}

\vspace{0.5cm}

{\em
\vspace{0.5cm}
 Institut f\"ur Theoretische Physik  \\ 
Leibniz Universit\"at Hannover \\
Appelstra\ss{}e 2, 30167 Hannover, Germany \\
\vspace{0.5cm}
 Max-Planck-Institut f\"u{}r Gravitationsphysik (Albert-Einstein-Institut)  \\
Am M\"u{}hlenberg 1, DE-14476 Potsdam, Germany }
\\[1cm]
 \texttt{ lorenzo.casarin@\{itp.uni-hannover.de, aei.mpg.de\}} 

\end{center}
\vspace{0.9cm}
%
\begin{abstract}
\noindent 
We compute the conformal anomalies for some higher-derivative (non-unitary) 6d Weyl invariant theories using the heat-kernel expansion in the background-field method. 
To this aim we obtain  the general expression for the  Seeley-DeWitt coefficient \(b_6\) for four-derivative differential operators with  background curved   geometry and gauge fields, which  was known only in flat space so far. 
We consider four-derivative scalars and abelian vectors as well as three-derivative fermions, confirming the result of the literature obtained via indirect methods.
We generalise the vector case by including the curvature coupling \( FF \mathrm{Weyl}\). 
\end{abstract}
 
\newpage
\tableofcontents

\setcounter{footnote}{0}

\section{Introduction}

The calculation of conformal anomalies for six-dimensional theories has recently been of interest, also in the higher-derivative case (see \cite{Cordova:2015fha,Cordova:2015vwa,Beccaria:2015uta,Beccaria:2015ypa,Beccaria:2017dmw,Tseytlin:2013fca,Herzog:2013ed,Mukherjee:2021alj} and references therein).
In the context of conformal field theory, six-dimensional spacetime plays a very important role, as no interacting unitary supersymmetric conformal field theory can exist in more than six dimensions \cite{Nahm:1977tg} and no example is known  even in the non-supersymmetric case. It is however difficult to study unitary theories in six dimensions due to the lack of perturbative renormalizability for standard 2-derivative actions. Higher-derivative theories, despite being non-unitary, can be considered as formal UV completion of standard 2-derivative theories \cite{Fradkin:1982kf} and can therefore help to shed light on the properties of conformal field theories and of the space of QFTs in higher dimensions, see  e.g.\ \cite{Ivanov:2005qf,Buchbinder:2016url,Buchbinder:2018lbd,Buchbinder:2020ovf}.

The conformal anomaly \(\mathscr A\) in 6 dimensions takes the form \cite{Bonora:1985cq,Deser:1993yx,Bastianelli:2000hi}
\begin{equation}\label{aaa}
\mathscr{ A} \cdot (4\pi)^3
=  g^{mn}\braket{T_{mn}}   \cdot (4\pi)^3
= - a \, \mathbb{E}_6 + c_1 \,  I_1 + c_2 \,  I_2 + c_3 \, I_3
\,,
\end{equation}
where \(\mathbb E_6\) is the  Euler density in \(6\) dimensions and the invariants \(I_i\) are built from the Weyl tensor (\(I_{1},I_2\sim\mathrm{Weyl}^3\), \(I_3\sim \mathrm{Weyl}\cd^2\mathrm{Weyl} \) -- see appendix~\ref{appA} for explicit expressions). \eqref{aaa} also appears in the UV divergent part of the effective action, and the anomaly coefficients \(a,c_i\) enter in the stress tensor two-, three- and four-point functions. In \eqref{aaa} we ignored scheme-dependent  total-derivative contributions. 

An efficient way of determining the  UV divergent part of the effective action, and equivalently the conformal anomaly coefficients, is the heat kernel method. By providing   a representation of the determinant of a differential operator preserving background covariance, the heat kernel is particularly suited to study 1-loop effects.
In the present case the  relevant terms are captured by
\begin{equation}\label{iab} 
\Gamma_\infty 
=  - \frac{   \log   \Lambda  }{(4\pi)^{3}}
 \int \!  \sqrt{g }  \ b_6 
 \,,
 \qquad     
 \mathscr{ A} = \frac{1}{(4\pi)^3} \,  b_6\,,
 \qquad       
 b_6 = b_6(\Delta_{\text{b}})-b_6(\Delta_{\text{f}}) \pm b_6(\Delta_{\text{gh}})\,,
\end{equation}
where \(b_6\) is a combination of the heat kernel coefficients  \( b_6(\Delta) \) of the operators \( \Delta \)  governing the quadratic fluctuations.  In writing \eqref{iab} we assumed  real bosons (b) and Weyl or Majorana fermions (f) in gamma-matrix representation.  The last term schematically represents ghost (gh) contributions.
The heat-kernel coefficients for second-order differential operators have been known for a long time and have been widely applied to physics \cite{Gilkey:1975iq,Avramidi:2000bm,Vassilevich:2003xt,Barvinsky:1985an}. The coefficients for higher powers of the Laplacian and its deformations have also been considered, albeit with less completeness, see e.g.\ \cite{Gilkey:1980,Fradkin:1981iu,Bugini:2018def,Avramidi:2000bm,Casarin:2019aqw}.  

In particular, for the scope of this paper we need to consider   operators of the form
\begin{equation}\label{iad}
\Delta_4 
=
	\cd^4
	+ V^{ mn } \cd_m \cd_n
	+ 2 N^{ m } \cd_m 
	+ U\,,
\end{equation}
where   \(V_{mn}=V_{nm}\), the covariant derivative contains spacetime as well as gauge connections  and the coefficient functions \(V,N,U\) are generally matrix-valued.
The coefficient \(b_6(\Delta_4)\) was recently computed in flat spacetime in \cite{Casarin:2019aqw} (see also \cite{Casarin:2021fgd})   using an  argument based on  special factorised cases in terms of two-derivative operators  first proposed by \cite{Fradkin:1981iu} in the context of four-dimensional quadratic gravity. Here we extend the result to include a geometrical background,
 thereby providing a direct way to compute the conformal anomaly coefficients.

We then use the newly obtained coefficient \(b_6(\Delta_4)\) to provide a direct calculation of the anomaly coefficients of some classically Weyl-invariant scalar, spinor and vector models.  Most of these   have been recently computed using indirect techniques \cite{Bugini:2018def,Beccaria:2017dmw,Beccaria:2015uta}; our results provide an independent confirmation based on a conceptually straightforward and well-established procedure. In the case of the vector we furthermore extend  the case studied in the literature by including an extra coupling with the background geometry with the structure  \(FF\mathrm{Weyl}\).

The fields considered in this paper also appear  as lower-spin contributions to   six-dimensional \((2,0)\) conformal supergravity theory,  where the graviton kinetic term is a combination of the \(I_i\)'s above.   This theory, constructed  in \cite{Butter:2017jqu} to a level which is sufficient for the one-loop  anomaly calculation, contains however a six-derivative operator which therefore escapes the scope of the present paper. It would be interesting to compute the conformal anomaly of this theory, for which the \(a\)-coefficients are known from holographic considerations \cite{Beccaria:2015uta} and suggest that \((2,0)\) conformal supergravity coupled to 26 \((2,0)\) tensor multiplets is anomaly free.

Finally, let us comment on  zero modes, which are ignored in this paper.  In general, a  differential operators like the one in \eqref{iad}  admits normalisable zero modes, which  have to be treated separately from the rest of the spectrum. 
For fourth-order operators that factorise into second-order ones exhibiting zero modes,  these will be naturally inherited, however there could be additional ones depending on the interplay between the two factors. These aspects fall outside the scope of this paper and  deserve further study.

This paper is organised as follows.
Section~2 presents some relevant facts about the heat kernel expansion and discusses the derivation of the heat-kernel coefficient \(b_6(\Delta_4)\)  using the factorisation Ansatz. In section~3 we apply such newly computed coefficient to the calculation of conformal anomalies \eqref{aaa} to four-derivative scalar, four-derivative gauge vector and three-derivative spinor. Appendix~A summarises notation. Appendix~B presents some more complete facts on the heat-kernel expansion that are useful for this paper and provides additional explicit formulae. 
Appendix~\ref{app::basis2} for completeness lists a basis for \(b_6(\Delta_4)\) used in one of the decompositions. 
Appendix~\ref{app::diagr} discusses some diagrammatic checks for our result of \(b_6(\Delta_4)\).

\section{Heat kernel coefficient \texorpdfstring{$b_6(\Delta_4)$}{b6(D4)}  on a geometric background  }

\subsection{Preliminary considerations}
\label{sec::hk}

Here we recall some basic facts about the heat-kernel expansion relevant for the calculation. Further  detail is given in appendix~\ref{app:HK}.

We consider  an elliptic   differential operator $\Delta  $ of   even order $2n$ defined on a $d$-dimensional  manifold without boundaries with the schematic structure
\begin{equation}\label{iaa}
\Delta = (- \cd^2)^{n} + \text{lower derivative terms},
\end{equation}
where \(\cd=\partial+\Gamma +A \) is a covariant derivative with geometric and gauge connection.
We denote the associated spacetime and internal curvatures as \([\cd,\cd]=R+\WW\).
One can express the logarithmically divergent part of \(\det \Delta\) as (see e.g.~\cite{Fradkin:1982kf,Avramidi:2000bm,Casarin:2021fgd} and references therein)
\begin{align}\label{aua} 
{(\log \det \Delta)_\infty }
=
- \frac{2 \log  \Lambda }{(4\pi)^{d/2}}
 \int \!  \sqrt{g } \ b_d(\Delta)
 \,,
\end{align}
where  \(\Lambda\) is the UV cutoff  and \(b_p \) is the trace of a local covariant quantity built using the differential operator as well as the covariant derivative and  it is defined modulo boundary terms. We shall refer to these coefficients as Seeley-DeWitt heat-kernel coefficients. Let us now consider two differential operators \(\Delta\) and \(\Delta'\).
Using the factorisation property of the determinant for combined operators we   obtain the key relation
\begin{align}\label{aub2}
 b_d(\Delta\Delta')= b_d(\Delta)+  b_d(\Delta') \,,
\end{align}
which allows one to relate the heat kernel coefficients of operators of different order (again modulo total derivatives).
We stress here that the relation \eqref{aub2}   is  valid for the \emph{logarithmic part} only and not for power-law divergences.

In the case of	 the  2-derivative operator
\begin{align}
\label{kah}
& 	\Delta_2 = -\cd^2 + X\,,
\end{align}
where     $\cd_m$  has internal and spacetime connections, 
 in \(6d\)
one  has
the expression \eqref{app2}  \cite{Gilkey:1975iq,Avramidi:2000bm,Vassilevich:2003xt}, which can be schematically represented as
\begin{align} \label{avb}
\begin{aligned}
	b_6(\Delta_2)
&= {} 
	b_6^\text{g}(\Delta_2) 
	+
	b_6^\text{gc}(\Delta_2)
	+
	b_6^\text{m}(\Delta_2)\,,
\end{aligned}
\end{align}
where we distinguished the purely gravitational terms (`g'), those that originate from the generally-covariantized flat-spacetime expression (`gc'), and the terms which mix gravitational and gauge terms (`m'), so that   \( b_6^\text{m}(\Delta_2)\)   vanishes in flat spacetime as well as \( b_6^\text{g}(\Delta_2) = \mathfrak E \cdot \tr \1\) (\(\mathfrak{E}\) is given in \eqref{geom} and \(\1\) is the identity in the internal space where tr acts).

\subsection{Derivation of \texorpdfstring{\(b_6(\Delta_4)\)}{b6(D4)} }

We are interested in the coefficient \(b_6(\Delta_4)\), where the operator \(\Delta_4\) has the structure \eqref{iad}. 
One can equivalently present the operator \eqref{iad} in the  `symmetric' form \eqref{zadd}.
Following \eqref{avb} we correspondingly decompose
\begin{equation}\label{aad}
\begin{aligned}
	b_6(\Delta_4) = 	b_6^\text{g}(\Delta_4) 
		+
		b_6^\text{gc}(\Delta_4)
		+
		b_6^\text{m}(\Delta_4)\,.
\end{aligned}
\end{equation} 
The strategy to compute it is the following. First, we make an Ansatz based on dimensional and covariance considerations, taking into account algebraic relations between different terms due to   Bianchi identities, symmetries of tensors and boundary terms. Then we consider special cases for \(\Delta_4\), where it can be decomposed as the produce of second-order  operators, \(\Delta_4 = \Delta_2\,  \Delta_2'\). Using  \eqref{aub2} with the explicit expression for \eqref{avb} in \eqref{aub}  allows us to gather enough information to reconstruct  \(b_6(\Delta_4)\). 

From these considerations it is immediate to see that 
\begin{equation}
b_6^\text{g}(\Delta_4)  =2	\, b_6^\text{g}(\Delta_2)
= 2 \mathfrak E \cdot \tr \1
 \,.
\end{equation}
Furthermore, this procedure was already applied to \eqref{iad} in \cite{Casarin:2019aqw} (see also \cite{Casarin:2021fgd}) to the flat-spacetime case, therefore 	\(b_6^\text{gc}(\Delta_4)\) can be immediately obtained,   
\begin{equation}\label{avdd}
\begin{aligned}
b_6^\text{gc}(\Delta_4) =& \tr \Big[ 
		 - \frac{1}{30}  \left( \cd_m \WW_{mn}  \right)^2
		+ \frac{1}{45} \WW_{mn} \WW_{nr} \WW_{rm}
\\
& \qquad
		+ \frac{1}{360} V_{mn} V_{nr} V_{rm}
		+ \frac{1}{480} V_{mn} V_{mn} V
		+\frac{1}{2\,880}V V V
		+ \frac{1}{30}V_{mn} \cd_{(n} \cd_{r)} V_{rm} 
\\
& \qquad
		+\frac{1}{120} V_{mn} \cd^2 V_{mn}
		 - \frac{1}{40} V_{mn} \cd_{m} \cd_{n} V
		+ \frac{1}{240} V \cd^2 V
		- \frac{1}{12}  V_{mn} V_{nr}  \WW_{mr} 
	\\
& \qquad
		+  \frac{1}{6} \WW_{mn} \cd_{(m} \cd_{r)} V_{rn}
		+  \frac{1}{24} V \WW_{mn} \WW_{mn}
		- \frac{1}{6}  V_{mn} \WW_{mr} \WW_{nr}  
\\
& \qquad
		- \frac{1}{3} \WW_{mn} \cd_{m} N_n 
		- \frac{1}{6}  V_{mn} \cd_{m} N_n 
		+ \frac{1}{12} V \cd_m N_m
		- \frac{1}{6}  N_m N_m
		 - \frac{1}{12}   U V
		 \Big],
\end{aligned}
\end{equation}  
where $V   = g^{mn}V_{mn}$ and \(\WW\) is the internal curvature.
What remains to be determined is therefore only 	\(b_6^\text{m}(\Delta_4)\).
On dimensional and covariance grounds we make the Ansatz 
\begin{equation}\label{avd}
\begin{aligned}
b_6^\text{m}(\Delta_4) =& \tr \big[
			c_1 R \WW_{mn} \WW_{mn} 
			+ c_2  R V_{mn} V_{mn}
			+ c_3  R V V
			+ c_4 RU
			+ c_5  R \cd^2 V
			+ c_6  R \cd_m \cd_n V_{mn}
			\\
			&  \qquad
			+ c_{7} R \cd_m N_m
			+ c_{8} R_{mn} V_{nr}V_{rm}
			+ c_{9} R_{mn} V_{mn}V 
			+ c_{10}  R_{mn} \cd^2 V_{mn}
			+ c_{11} R_{mn} V_{nr} \WW_{mr}
						\\
						&  \qquad
			+ c_{12} R_{mn} \WW_{mr} \WW_{nr}
			+ c_{13}  R_{mnrq} V_{mr} V_{nq} 
			+ c_{14}  R_{mnrq} \WW_{mn} \WW_{rq} 
			+ c_{ 15 }   R_{mnrq} R_{mnrq}  V   
			\\
			&  \qquad
			+ c_{ 16 }     R_{mrqk} R_{nrqk}  V_{mn}
			+ c_{ 17 }    R_{mn} R_{mn}  V 
			+ c_{ 18 }    R^2  V 
			+ c_{ 19 }     R_{mn} R_{mr}  V_{rn} 
			\\
			&   \qquad
			+ c_{ 20 }     R R_{mn}  V_{mn} 
			+ c_{ 21 }     R_{mnrq} R_{mr}  V_{nq} 
			\big]
			\,.
\end{aligned}
\end{equation}  
All other combinations  vanish or reduce to these by means of the Bianchi identities, integration by parts and symmetry properties. 
As we shall explain, we find
\begin{equation}\label{ccs}
\begin{aligned} 
c_1 &= \frac 1 {36}    \,,      &
c_2 &= \frac 1 {144}  \,,       &
c_3 &= \frac 1 {288}   \,,      &
c_4 &= -\frac 1 6     \,,    &
c_5 &= \frac 1 {60}    \,,          &
c_6 &= - \frac 1 {20}  \,,       &
c_7 &= \frac 1 6    \,,       \\
c_8 &= -\frac 1 {36}  \,,        &
c_9 &= -\frac 1 {72}   \,,      &
c_{10} &= -\frac 1 {60}  \,,        &
c_{11} &= -\frac 1 {12}    \,,     &
c_{12} &= \frac 1 {45}   \,,      &
c_{13} &= \frac 1 {36}    \,,     &
c_{14} &= \frac 1 {90}     \,,    \\
c_{15} &= \frac 1 {360}   \,,      &
c_{16}&= -\frac 1 {90}    \,,     &
c_{17} &= -\frac 1{ 360}   \,,       &
c_{18} &= \frac 1 {144}    \,,     &
c_{19} &= \frac 1 {45}    \,,     &
c_{20} &= - \frac 1 {36}  \,,       &
c_{21} &= -\frac 1 {90}    \,.
\end{aligned}
\end{equation}
The full   explicit expression of \(b_6(\Delta_6)\) is given in \eqref{avdd2}.

To fix the values of the coefficients \(c_i\)'s we resort to the following two  decompositions.
\begin{enumerate}
\item \(\Delta_4 = \Delta^X_2 \Delta^Y_2 \),  where the 2-derivative operators have the structure \eqref{kah} and   the background gauge connection is non-abelian. 

\item \(\Delta_4 = \Delta_+ \Delta_-\), with an abelian gauge connection  and  \( \Delta_\pm = - (\cd_m \pm B_m)^2  \). 
\end{enumerate} 
In total we find  an overdetermined system of   49  equations with   unique solution \eqref{ccs}.
The following two subsections provide details of the derivation.

\subsubsection{Decomposition 1}
The fourth-order operators obtained from the composition 
\begin{equation}\label{aag}
\Delta_4 = \Delta^X_2 \Delta^Y_2
\end{equation}
has the structure \eqref{iad} with $V_{mn} = - \delta_{mn} (X+Y) $, $N_m = - \cd_m Y$ , $ U = XY- \cd^2 Y $, and therefore  $  V = - 6 (X+Y) $. 

From the general expression \eqref{avd} we get
\begin{equation}\label{dec1}
\begin{aligned}
b_6^\text{m}(\Delta_4) 
= &\tr \Big[ 
  c_{1}  \WW_{m n  } \WW_{m n  } R
  + c_{12}  \WW_{m  r  } \WW_{m n  } R_{n r} 
+  c_{14}  \WW_{m n  } \WW_{r s   } R_{m n r s }
\\&\qquad
- (  6 c_{18}+c_{20}) R^2 X   
- (   6 c_{18} +  c_{20}) R^2 Y 
\\&\qquad
- (   6 c_{17} +  c_{19}+  c_{21}) R_{m n} R_{m n} X 
- (  6 c_{17} + c_{19} + c_{21}) R_{m n} R_{m n} Y
\\&\qquad
+ (    6 c_{2}+ 36 c_{3}+ c_{8}+ 6 c_{9} +c_{13}) R X ^2
+ (    6 c_{2}+ 36 c_{3}+ c_{8}+ 6 c_{9} +c_{13} ) R Y^2
\\&\qquad
- (   6 c_{15} +  c_{16}) R_{m n r s } R_{m n r s } Y  
- (   6 c_{15} +  c_{16}) R_{m n r s } R_{m n r s } X 
\\&\qquad
- (  6 c_{5}+ c_{6} + c_{10} ) R \cd^2X 
- (     c_{4}+ 6 c_{5}+  c_{6}+  c_{7} + c_{10} ) R \cd^2Y 
\\&\qquad
+ ( 12 c_{2}+ 72 c_{3}+ c_{4}+ 2 c_{8}+ 12 c_{9} +2 c_{13}) R X Y  
 \Big] 
\,.
\end{aligned}
\end{equation}
and from the factorisation we have
\begin{equation}\label{dec11}
\begin{aligned}
&b_6^\text{m}(\Delta_2) +b_6^\text{m}(\Delta_2') 
\\ &\qquad=
 \tr \Big[ 
- \frac{1}{36}   \WW_{m n  } \WW_{m n  } R
- \frac{1}{45}   \WW_{m  r  } \WW_{m n  } R_{n r} 
- \frac{1}{90}   \WW_{m n  } \WW_{r s   } R_{m n r s } 
+ \frac{1}{72} R^2 X
+ \frac{1}{72}  R^2 Y 
\\&\qquad\qquad\qquad
- \frac{1}{180}  R_{m n} R_{m n} X 
- \frac{1}{180}  R_{m n} R_{m n} Y  
- \frac{1}{12} R X ^2
- \frac{1}{12}   R Y ^2 
\\&\qquad\qquad\qquad	
+ \frac{1}{180}  R_{m n r s } R_{m n r s } X
+ \frac{1}{180}  R_{m n r s } R_{m n r s } Y   
+ \frac{1}{30}  R \cd^2X 
+ \frac{1}{30}   R \cd^2 Y 
  \Big]
\,.
\end{aligned}
\end{equation}
Equating the two we obtain 14 linear equations.

\subsubsection{Decomposition 2}

We are considering
\begin{equation}\label{aaz}
\Delta_4 = \Delta_+ \Delta_- ,
\hspace{3em}
- \Delta_\pm = (\cd_m \pm B_m)^2 = \cd^2 \pm 2 B_m \cd_m  \pm (\cd_m B_m) + B_m B_m
\end{equation}
with \( \cd = \partial+ \Gamma +A  \), \(A\) being an abelian connection. 
The field strengths therefore read
\begin{equation}\label{aai}
\WW^{\pm}_{mn} \equiv [\cd^\pm_m,\cd^\pm_n] = \WW_{mn} + [B_m, B_n] \pm (\cd_m B_n - \cd_n B_m)\,.
\end{equation} 
The coefficients for the operator $\Delta_4$ \eqref{aaz} in the notation \eqref{iad} read
\begin{align}
\label{aba}
V_{mn} & = - 4 \cd_{(m} B_{n)} + 2 B^2 \delta_{mn} - 4 B_{(m} B_{n)}  \,,
\qquad
\qquad 
V  =   -4 \cd_n B_n  +8 B^2\,,
\\
\label{abb}
N_m & = 
- B_n R_{mn}
- \cd^2 B_m - \cd_m \cd_n B_n + \cd_m B^2 + B_m B^2
\\* \nonumber
& \qquad - B^2 B_m - 2 B_n \cd_n B_m - B_m \cd_n B_n + 2 B_n \WW_{nm}  \,,
\\
\label{abc}
U & = - \cd^2 \cd_n B_n + \cd^2 B^2 - 2 B_m \cd_m \cd_n B_n + 2 B_m \cd_m B^2 - (\cd_n B_n)^2 + B^4
\\* \nonumber
&\qquad  + (\cd_n B_n)B^2 - B^2 \cd_n B_n -2 \cd_m B_n \WW_{mn} - 2 B_m B_n \WW_{mn}  + 2 B_m \cd_n \WW_{mn}\,.
\end{align} 
The only place where the spacetime curvature explicitly appears is  \(B_n R_{mn}\) in \(N_m\).

In considering the factorisation Ansatz \eqref{aub2},  for simplicity we    focus on terms   of order 1, 2, 4, 5 and 6 in \(B\). It turns out that this provides us with enough information to determine \(b_6^\text m(\Delta_4)\).

To start, we need to determine a basis for the invariants that can appear in the expression of \(b_6(\Delta_4)\). In doing so one needs to be careful about the possibility of adding total derivatives, the symmetries of the objects involved and their algebraic relations. 
We identified a basis consisting of 45 elements listed in appendix~\ref{app::basis2}.
In such a basis, we can evaluate \eqref{avd} as
\begin{equation}\label{dec2}
\begin{aligned}
b_6^\text{m}(\Delta_4) 
& = 
c_{12}  \WW_{a n} \WW_{a m} R_{m n} 
+ 4 c_{11}   \WW_{an} B_{a } B_{m} R_{m n} 
- 16 (  c_{13}  + 2 c_{9} ) B^2B_{m} B_{n} R_{m n} 
\\&\quad
-4 ( 2 c_{13} + c_{19} ) B_{a } B_{m} R_{a n} R_{m n} 
+ 2 (4 c_{17}  + c_{19}  + c_{21} ) B^2R_{m n} R_{m n} 
+  c_{1}   \WW_{a m} \WW_{a m} R 
\\&\quad
+ (  24 c_{2}  + 64 c_{3}  + c_{4}  + 4 c_{8}  + 16 c_{9}+4 c_{13} ) B^2B_{m} B_{m} R
- 8 (2 c_{8} -   3 c_{13}  ) B_{a } R_{m n} \cd_{a }\cd_{n}B_{m}
\\&\quad
- 2 (  2 c_{6}  + 	 c_{7} -  2 c_8  + 2 c_{13}  + 2 c_{20}    ) B_{a } B_{ m} R_{a m} R 
- 4 (  c_{10}  -  2 c_{17}  -  c_{21} ) B_{a } R_{m n} \cd_{a }R_{m n}
\\&\quad
+ 2 (4 c_{18}  + c_{20} ) B^2R^2
+  c_{14}  \WW_{a m} \WW_{n c } R_{a m n c }
- 4 (   2 c_{8} -  2 c_{13}  + c_{21}   ) B_{a } B_{m} R_{n c } R_{a n m c }
\\&\quad
+ 4 (  2 c_{13}  - c_{16} ) B_{a } B_{m} R_{a n c r} R_{m n c r}
+ 2 (4 c_{15}  + c_{16} ) B^2R_{m n c r} R_{m n c r}
\\& \quad
+ 2 (4 c_{18}  + c_{20} ) B_{a } R \cd_{a }R 
+ 2 (4 c_{15}  + c_{16} ) B_{a } R_{m n c r} \cd_{a }R_{m n c r}
\\&\quad
-2  (  c_{4} + 4 c_{6} + 2 c_{7} - 2 c_{8} - 4 c_{9} - 2 c_{13}   ) B_{a } R \cd_{a }\cd_{m}B_{m}
\\&\quad
+ 4(  c_{8}  - 2 c_{10}  - 2 c_{13}     ) R_{a n} \cd_{m}B_{n} \cd_{m}B_{a } 
+ 8 c_{13} B_{a } R_{a c m n} \cd_{c }\cd_{n}B_{m}
\\&\quad
+ (     16 c_{3}  -  c_{4}  - 4 c_{6}  - 2 c_{7}  + 8 c_{9}  + 8 c_{13}  ) R \cd_{a }B_{a } \cd_{m}B_{m} 
\\&\quad
+ 2 (    c_{4}  + 8 c_{5}  + 2 c_{6}  +  c_{7}  +   c_{8}  + 2 c_{10} ) B_{a } R \cd^2B_{a } 
- 8 (   2 c_{9}  +  3 c_{13} ) B_{a } R_{a n} \cd_{n}\cd_{m}B_{m} 
\\&\quad
- (   c_{4} +   4 c_{5} + 4 c_{6} +2 c_{7}  +  2 c_{10}  )  R \cd^2\cd_{a }B_{a } 
-8 (   c_{10} + c_{13} ) B_{a } R_{a m} \cd_{n}\cd_{n}B_{m} 
\\&\quad
+ 2(   4 c_{2}   - 2c_{6}  -   c_{7}  + 2 c_{8} - 2 c_{13}   ) R \cd_{a }B_{m} \cd_{m}B_{a }
\\&\quad
+ 2( 4 c_{2}  +  c_{4}  +8 c_{5}  + 2 c_{6}  +  c_{7}  +  c_{8}  + 2 c_{10}  ) R \cd_{m}B_{a } \cd_{m}B_{a }
\\&\quad
-(  2  c_{4}  +2 c_{7}  +   c_{11}  ) \WW_{a m} B_{a } \cd_{m}R 
- 4 c_{8}  B_{a } R_{m n} \cd_{n}\cd_{m}B_{a } 
\\&\quad
+ 2 (  2 c_{6}  + c_{7} + 2 c_{10}  + c_{19}  + 2 c_{20}  ) B_{a } R_{a m} \cd_{m}R 
- 2 c_{11}  B_{a } R_{m n} \cd_{n}\WW_{a m} 
\\&\quad
+   2 c_{11}   B_{a } R_{a m} \cd_{n}\WW_{mn} 
+ 4 (2 c_{10}  - 2 c_{16} -  c_{21} )  B_{a } R_{a m n c } \cd_{c }R_{m n} 
\\&\quad
+  2 c_{11}   \WW_{m n} B_{a } \cd_{n}R_{a m} 
+ 4 (2 c_{10}  + c_{19}  -  c_{21} ) B_{a } R_{m n} \cd_{n}R_{a m} 
\end{aligned}
\end{equation}
and from the factorisation we have
\begin{equation}\label{dec22}
\begin{aligned}
&b_6^\text{m}(\Delta_+) +b_6^\text{m}(\Delta_-) 
\\ & \qquad
=
- \frac{1}{45}    \WW_{a n} \WW_{a m} R_{m n} 
+   \frac{1}{3}   \WW_{an} B_{a } B_{m} R_{m n}  
+ \frac{14}{45}  B_{a } B_{m} R_{a n} R_{m n}  
- \frac{1}{36}   \WW_{a m} \WW_{a m} R \\& \qquad\qquad
+ \frac{11}{45}    B_{a } B_{ m} R_{a m} R  
- \frac{1}{90}  \WW_{a m} \WW_{n c } R_{a m n c }
- \frac{22}{45}    B_{a } B_{m} R_{n c } R_{a n m c } \\&\qquad\qquad
-   \frac{4}{15}    B_{a } B_{m} R_{a n c r} R_{m n c r} 
+ \frac{2}{45}   B_{a } R \cd_{a }\cd_{m}B_{m} 
- \frac{8}{9}    B_{a } R_{m n} \cd_{a }\cd_{n}B_{m} 
- \frac{1}{5}   R \cd_{a }B_{a } \cd_{m}B_{m} \\&\qquad\qquad
+ \frac{1}{18}   B_{a } R \cd_{m}\cd_{m}B_{a }    
+ \frac{3}{10}    R \cd_{a }B_{m} \cd_{m}B_{a } 
+ \frac{1}{5}   R_{a n} \cd_{m}B_{n} \cd_{m}B_{a } 
- \frac{1}{12}   \WW_{a m} B_{a } \cd_{m}R \\& \qquad\qquad
- \frac{1}{6}   B_{a } R_{m n} \cd_{n}\WW_{a m} 
+ \frac{1}{6}   B_{a } R_{a m} \cd_{n}\WW_{mn} 
+ \frac{1}{6}    \WW_{m n} B_{a } \cd_{n}R_{a m}  
+ \frac{4}{45}   B_{a } R_{a m} \cd_{n}\cd_{n}B_{m} \\&\qquad\qquad
+ \frac{4}{9}  B_{a } R_{a n} \cd_{n}\cd_{m}B_{m} 
- \frac{1}{9}   B_{a } R_{m n} \cd_{n}\cd_{m}B_{a }  
- \frac{2}{9} B_{a } R_{a c m n} \cd_{c }\cd_{n}B_{m}
\,.
\end{aligned}
\end{equation}
Equating the two expressions we obtain  35 linear equations.

\section{Applications}

\subsection{Four-derivative scalar field} 
 
 A four-derivative Weyl-covariant differential operator  \cite{Beccaria:2015uta,Beccaria:2017dmw}  in \(d\)-dimensions was constructed by Paneitz (cf.~\cite{Paneitz_2008}; the \(4d\) case was first given in \cite{Fradkin:1981jc} and \cite{Riegert:1984kt}),
 \begin{equation}
 \label{bai}
\Delta_4 = \cd^4 + \cd_m[(4 S_{mn} - (d-2) g_{mn} S  )\cd_n]
- (d-4) S_{mn} S_{mn} + d \frac{d-4}{4}   S^2 - \frac{d-2}{2} (\cd^2 S )
\end{equation}
where \(S_{mn} \) is the Schouten tensor
\begin{equation}\label{sch}
S_{mn} = \frac{1}{d-2}  \Big[  R_{mn} - \frac 1{2 (d-1) }  R g_{mn} \Big],
\qquad\qquad
S = S_{mm} = \frac{1}{2(d-1)} R
\,.
\end{equation}
Such operator allows one to consider the following Weyl-invariant action in \(6d\) for a real scalar, from which we can  compute the corresponding effective action and conformal anomaly via \eqref{iab},
\begin{equation}\label{ths}
S = 
\frac12 \int \dd{^6x  \sqrt{g} }    \phi \Delta_4 \phi 
\,,
\qquad\qquad
b_6 = b_6(\Delta_4)\,.
\end{equation}
The operator \eqref{bai} is written in the symmetric  form \eqref{zadd}. Direct application of \eqref{avdd2}  gives
\begin{equation}\label{baj}
(a,c_i) 
 = \frac{1}{7!} \left( \frac{4}{9} ,  \frac{224}{3} , 8,- 10  \right)
,
\end{equation} 
in agreement with   the recent independent analysis of  \cite[(16)-(19) with \(k=2\)]{Bugini:2018def}.

\subsection{Four-derivative gauge vector}
  
We consider the following Weyl-invariant action for an abelian gauge vector \(A_m\),
\begin{equation}\label{raa}
\begin{aligned}
S&  = \int \!  \sqrt g \   \left[
\cd_r F_{rm} \cd_n F_{nm} - \left(R_{mn}    -\frac15 g_{mn} R \right)F_{mp} F_{np}  
\right]  
+  \, \xi
\int \!  \sqrt g      F_{mn } F_{rs} W_{mnrs} 
\end{aligned}
\end{equation}
where  \(F_{mn} = \cd_m A_n - \cd_n A_m\) is the field strength. The first integral provides a Weyl-invariant kinetic term for \(A_m\) as considered in \cite{Beccaria:2017dmw}. The second integral  is Weyl-invariant by itself (\(W_{\cdot\cdot\cdot\cdot}\) being the Weyl tensor) and    can therefore be added with an  arbitrary numerical coefficient \(\xi\). 
In terms of the gauge field \(A_m\) the action reads
\begin{equation}\label{raa2}
S =  
\frac1 {2 } \int \! \sqrt g \    A_m   [\Delta_{4A} ]_{mn}   A_n 
+ \frac1 {2 }   \int\! \sqrt g \  A_m    \cd_m \cd^2 \cd_n A_n
\,,
\end{equation}
where the  operator \( \Delta_{4A} \) is a four-derivative one. It is more convenient to present it in the symmetrised form \eqref{zadd} with coefficients 
\begin{align}
\nonumber
	[\hat V_{mn}]_{ac} &=
(1+\xi) g_{ac} R_{mn} 
-(1-\xi)g_{mn} R_{ac} 
-   \frac{2+\xi}{5}   g_{ac} g_{mn} R 
\\ \label{rab} 
& 
\qquad\quad
+ \frac{2+\xi}{5}   R g_{m (a}g_{c)n}
 -  2 (1+\xi)  g^{ (m   }_{ (a}R_{c)} ^{n)}  
 +4 \xi R_{ a(mn)c } 
 \,,
\\ 
[\hat N_m]_{ac}&=
\frac{1 -3 \xi}{2}  \cd _{(a}R_{c)m} 
+ \frac{1 + 3 \xi}{20} g_{m (a}  \cd _{c)}R 
\\
[\hat U]_{ac}&=
 \frac{1  - \xi }{2} R_{am} R_{cm} 
 - \frac{1+\xi}{2} R_{ m s } R_{amcs}
+ \frac{ 2 + \xi }{10} R_{ac} R 
- \cd ^2 R_{ac}
+ 2  \xi  R_{ a m r s } R_{ c r m s }\,.
\end{align}
The second term in \eqref{raa2} can be gauge-fixed away by choosing the covariant gauge \(\cd_m A_m =0\) and averaging over gauges with the Gau\ss{}ian weight \(-\cd^2\).

The effective action for \eqref{raa} thus constructed reads  
\begin{equation}\label{rba}
Z= \left[  \frac{\det \Delta_{4A} }{   [ \det  \Delta_{2,0} ] ^3  }  \right]^{1/2} \,,
\qquad\qquad
\Delta_{2,0}  = - \cd^2 \,,
\end{equation}
where \(\Delta_{2,0}\) acts on scalars and comes from ghost and gauge fixing contributions.
The divergent part of the  effective action has therefore the structure \eqref{iab}
governed by the  coefficient
\begin{equation}\label{rbb}
b_6 = b_6( \Delta_{4A}  )- 3  b_6(-\cd^2  )\,,
\end{equation}
where the first term can be evaluated with \eqref{avdd2} and the second one with \eqref{app2}.
In computing \(  b_6(\Delta_4)  \) we use that \(\1 \) is the identity in the space of  \(6\) dimensional vectors and that the curvature \(\WW \) is given by \([\WW_{mn}]_{ac}=  R_{mnac}   \).

In the form  \eqref{iab}  
we obtain
\begin{equation}
\begin{aligned}
a&= \frac{275}{8\cdot 7!}
\,,
&
c_1 &= \frac{28}{7!}   (  97  -  60 \xi +4 \xi^2- 4 \xi^3 ) \,,
\\
c_2& = \frac{1}{7!}   ( 911-840 \xi+392 \xi^2-392 \xi^3 )  
\,,\qquad
&
c_3& = - \frac{150}{7!}  \,.
\end{aligned}
\end{equation} 
The case \(\xi=0\) was considered in  \cite{Beccaria:2017dmw} via indirect methods; our result agrees.  We notice that \(\xi\) does not affect the \(a\) coefficient (as expected) and that \(c_1\) and \(c_2\) do not exhibit common zeroes. The fact that \(\xi\) does not enter \(c_3\) is probably accidental at one-loop.

\subsection{Three-derivative fermion}
We consider here a three-derivative Weyl spinor \(\Psi\) with the kinetic operator given in \cite{Beccaria:2017dmw} (see also \cite{Butter:2017jqu})
\begin{equation}\label{uba}
S = \int \bar \Psi \Delta_3 \Psi\,,
\qquad\qquad
-i\Delta_3 = \slashed \cd^3 + 2S_{mn}\gamma_m \cd_n + \gamma^m \cd_m S
\,,
\end{equation}
where \(S_{mn}\) is the Schouten tensor as in \eqref{sch}. In \eqref{uba} and in the following we consider Dirac gamma matrix notation with \(\{\gamma_m,\gamma_n\} = 2 g_{mn}\), \(\gamma_m\) being \(8\)-dimensional.

We have \eqref{iab} with
\begin{equation}\label{jedf}
b_6 = - b_6(\Delta_3) \equiv
b_6(\Delta_1) -b_6(  \Delta_3 \Delta_1)\,,
\qquad\qquad
\Delta_1  = i \slashed{\cd}\,,
\end{equation}
where we evaluate the heat kernel coefficient \(b_6(\Delta_3)\) considering composition \( \Delta_3 \Delta_1\) with the first order Dirac operator \(\Delta_1  \) (acting on Weyl spinors)  and applying \eqref{aub2}. 
The four-derivative   operator \( \Delta_3 \Delta_1\) has  the structure \eqref{iad} with\footnote{The self-adjoint requirements discussed in  section~\ref{sadj} are not to be imposed, as the operator \(\Delta_{3+1}\) does not come from functional integration.}
\begin{equation}\label{ubc}
\begin{aligned}
V_{mn}& =
2 \gamma_r \gamma_{ ( n  } S_{m ) r}
-\frac 12  R g_{mn}\,,
\qquad\qquad
N_m  = \frac12 \cd_a S\,   \gamma^a \gamma^m -\frac14 \cd_m R\,,
\\
U & =
		\frac 1 {16}R^2
		 -\frac14\cd^2 R 
		 + \frac14S_{ma}R_{anrs} \gamma^m\gamma^n\gamma^r\gamma^s\,,
\end{aligned}
\end{equation}
which  via \eqref{avdd2} results in 
\begin{equation}\label{ubp}   
b_6(  \Delta_3 \Delta_1  ) 
 =  \frac 1 {7!} \left[
- \frac{  10   }{  9   }  \mathbb E_{6} 
- \frac{  448   }{  3   }   I_1 
- \frac{   172  }{   3  }   I_{2}
+ 4 I_{3}
 \right]
 \,. 
\end{equation}
The heat kernel coefficient of the Dirac operator can be computed by squaring it and using \eqref{avdd2}, which gives
\begin{equation}\label{uaa} 
\begin{aligned}
b_6(\Delta_1) 
& = \frac 12 b_6\Big[ \left( \Delta_1 \right)^2\equiv -\cd^2 + \frac 14 R\Big]
\\
& =
\frac 1 {7!} \left[
 \frac {191}{144}  \mathbb E_{6} 
+\frac{448}3  I_1 
- 16  I_{2}
-20  I_{3}
 \right].
\end{aligned}
\end{equation}
The expression \eqref{uaa} was originally obtained in \cite{Bastianelli:2000hi}.
In computing \eqref{ubp} and \eqref{uaa} we have   used
\([\cd_m,\cd_n]= \frac14 R_{mnrs} \gamma^r \gamma^s\)
and 
\(R_{mnrs} \gamma^m\gamma^n\gamma^r\gamma^s = -2 R\).
Here \(\mathbbm 1\)  of \eqref{avdd2} is   the identity in the \( 8 \)-dimensional spinor space, where \(\tr\) is taken. 

In conclusion the conformal anomaly coefficients derived from \eqref{iab} with \eqref{jedf} are
\begin{equation}\label{d3a}
(a,c_i)
=
\frac{1}{7!} 
\left(
\frac{39}{16} ,
\frac{896}{3},
\frac{220}{3} ,
-24 
\right).
\end{equation}
Our result \eqref{d3a} coincides with that of   \cite{Beccaria:2017dmw}  obtained via indirect methods,\footnote{Our values \eqref{d3a}  are double those given in (6.3) of \cite{Beccaria:2017dmw}, since formal Majorana-Weyl spinors are used there} thereby providing an  independent  direct confirmation.

\subsection*{Acknowledgments}
 I thank Arkady Tseytlin for  many  discussions related to this project and for reading the  draft of this manuscript. I also thank Ivan Avramidi and Dmitri  Vassilevich for correspondence.
Many of the calculations presented in this paper have been carried out with the aid of the software Wolfram Mathematica and the xAct package suite \cite{xAct,Nutma:2013zea,DBLP:journals/corr/abs-0803-0862,Frob:2020gdh}.

\appendix

\section{Notation and conventions}
\label{appA}
We work in 6 euclidean dimensions and indicate spacetime indices with Latin lowercase letters.
The metric is \(g_{mn}\) and we do not distinguish between upper and lower indices.
We write the covariant derivative with Levi-Civita and gauge connection as
\(\cd = \partial + \Gamma + A\). 
We introduce the gauge and Riemann curvatures
\begin{equation}\label{opa}
\begin{gathered}
\WW_{mn}= \partial_m A_n - \partial_n A_m+[A_m,A_n]
\,,
\\
R\indices{^m_{nac}} = \partial_a \Gamma_{nc}^m \pm \cdots \,,
\qquad\qquad
R_{mn} = R\indices{^a_{man}}\,,
\qquad\qquad
R = R^m_m\,,
\end{gathered}
\end{equation}
so that \([\cd_m,\cd_n]\phi = \WW_{mn}\phi\) on a spacetime scalar and 
\([\cd_m,\cd_n]V_a = R_{mnac}V^c\) on a spacetime vector without gauge indices. We reserve \(F_{mn}\)  for the Maxwell field-strength and keep \(\WW_{mn}\) in the general case. 
The Weyl tensor reads
\begin{equation}\label{ugya}
W_{mnrs}
=
 R_{mnrs}
 +   g_{ [n } ^{  [r } R^{ s] }_{  m ]}   
 + \frac{1}{10} g_{m [ r} g_{s ] n} R \,.
\end{equation}

We note the following identities that are used in the main text without mentioning,
\begin{equation}\label{ids}
\begin{gathered}
2 (\cd_m F_{mn})^2 =  (\cd_m F_{rs})^2 
+  2 F_{ma}F_{na} R_{mn}   
 - R_{mnrs}  F^{mn} F^{rs}   
 +  \text{t.d.}, 
\\
2 R_{rnac} R_{macn} = - R_{rnac} R_{mnac} 
\end{gathered}
\end{equation}

Basis of the anomaly:
\begin{equation}\label{krt}
\begin{aligned}
\mathbb E_6  & 
= - \varepsilon_{mnrspq} \varepsilon_{abcdef}  R^{mnab}   R^{rscd}   R^{pqef}
=   -32 R_{mnac} R_{acsr} R_{srmn}	+\ldots\,,
  \\
I_1   &  = W_{a m n  c} W_{m r s n} W_{r a c s}\,,
\\
I_2  &   = W^{a c  m n} W^{r s a c} W^{m n r s}\,,
  \\
I_3   &  = 
W_{amnr}\Big[    
g_{ac}\cd^2
+ 4 R_{a c} 
- \frac{6}{5} g_{ac} R  
\Big] W_{c mnr}\,.
\end{aligned}
\end{equation}

\section{Relevant facts about the heat kernel }
\label{app:HK}

In this appendix we provide further details about the heat-kernel expansion to complete the discussion of section~\ref{sec::hk}.

\subsection{Generalities}

A standard representation for the determinant of a differential operator of order \(\ell\) is 
\begin{equation}\label{waa}
\log \det \Delta_\ell
= - \int  \dd{^dx} \sqrt{g} \intl_\varepsilon^\infty \frac{dt}{t}\,  \tr \braket{x |  e^{-t \Delta_\ell}   |x }\ , 
\end{equation}
where tr is the trace over internal indices of the operator and $\varepsilon = \Lambda^{-\ell}$ is a UV cutoff. The matrix element in the integrand is the heat kernel. It  has  an  asymptotic expansion for $t \to  0^+$    that allows us to write  (see e.g.~\cite{Barvinsky:1985an,Vassilevich:2003xt,Avramidi:2000bm,Casarin:2021fgd})
\begin{equation}\label{kac}
\tr  \braket{x | e^{-t \Delta_\ell} |x }
\equiv  \tr  K(t;x,x;\Delta_\ell) 
	\simeq \sum_{p\geq 0} \frac{2}{(4\pi)^{d/2}\,  \ell}\  t^{(p-d)/\ell} \ a_p(\Delta_\ell, d,g;x) \ .
\end{equation}
The  Seeley-DeWitt coefficients  $a_p$  are local covariant expressions of dimension $p$ 
constructed  out of the background  metric and gauge field, exhibiting an explicit  nontrivial dependence on the spacetime dimension $d$ when \(\ell \neq 2\), see e.g.~\cite{Gusynin:1988zt} for explicit examples. 
However, we are interested in the  spacetime integral  of the trace of the Seeley-DeWitt coefficients, which in the present context acquire the interpretation of a Lagrangian density.
We therefore focus on the simpler invariant quantity
\begin{equation}\label{kav}
b_p(\Delta_\ell, d,g;x) = 
\tr a_p(\Delta_\ell, d,g;x)  
\qquad\quad 
\text{modulo total derivatives},
\end{equation} 
which with abuse of notation we also call heat-kernel coefficients.
Some of the arguments of \(b_p\) are often omitted.
As a consequence, one can express the  divergent part of \eqref{waa} as 
\begin{equation}\label{kag}
\begin{split}
&{(\log \det \Delta_\ell)_\infty }
 =
    - \frac{2}{(4\pi)^{d/2}} \Big[
    \sum^{d-1}_{p=0} 
	\frac{{B_p(\Delta_\ell) }	}{d-p}  \Lambda^{d-p}
	+
	B_d(\Delta_\ell) \log \frac{\Lambda}{\mu} 
 \Big], 
 \\
 & B_p(\Delta_\ell)  = \int \dd{^d x \sqrt{g}   } b_p(\Delta_\ell)\ .
 \end{split}
 \end{equation}
where \(\mu \)  is a renormalization scale and the explicit dependence on \(\ell\) dropped out. In \eqref{kag} we find both power-law divergences (which we ignore for the scope of this paper) and the logarithmic term relevant for the conformal anomalies. 
In dimensional regularisation one has only the logarithmic term of \eqref{kag} with the formal substitution \(\log \frac{\Lambda}{\mu} \to \frac{1}{n-d} \), where \(n\) is the original integer number of dimensions. 

Let us now consider two differential operators \(\Delta\) and \(\Delta'\).
Using the factorisation property of the determinant for combined operators we further obtain the key relation
\begin{align}\label{aub}
 b_d(\Delta\Delta')= b_d(\Delta)+  b_d(\Delta') \,.
\end{align}
This factorization Ansatz is only true for the coefficient $b_{ p =d}$, i.e.\  with  the index $p$ equal to   the spacetime dimension $d$. 
Indeed, only the logarithmically divergent term of the expansion \eqref{kag} is universal, while the power-law  divergences are regularization-dependent.  This   implies that the factorisation Ansatz \eqref{aub} does not capture the explicit \(d\)-dependence of the \(b_p\)'s.

\subsection{Explicit formulae}
In this section we give some explicit expressions for the heat kernel coefficients of two- and four-derivative  elliptic operators.
We consider here operators of the standard forms
\begin{align}
\label{kad}
\Delta_2 &= - \cd^2 + X\,,
& 
	\Delta_4
 &=
 \cd^4
			+{ V }_{mn}  \cd_m    \cd_n
			+ 2 {N}_m \cd_m  
			+ {U}
			\,,
\end{align}
with \( V_{mn} = V_{nm}\); \(X,V,N,U\) are  covariant coefficient functions,   in general matrix-valued. In \eqref{kad}, \(\Delta_2 \) is the most general second order  operator and \(\Delta_4\) is  the most general  fourth-order   
operator without  3-derivative term. At this point, no further restriction is imposed on the coefficient functions, as the operators are not necessarily self-adjoint.

 Let us define the purely geometrical contribution as
\begin{equation}\label{geom}
\begin{aligned}
\mathfrak E 
={}
& 
		\frac{1}{945} R_{mn} R_{pq} R_{mpnq}
		+ \frac{1}{7\,560} R_{mn} R_{mpqr} R_{npqr}
		- \frac{4}{2\,835} R_{mn} R_{np} R_{pm}
\\
& \qquad
		+ \frac{17}{45\,360} R\indices{_{mn}^{pq}}  R\indices{_{pq}^{rs}}  R\indices{_{rs}^{mn}} 
		- \frac{1}{1\,620} R\indices{_m^p_n^q}  R\indices{_p^r_q^s}  R\indices{_r^m_s^n}  
		+ \frac{1}{840} R_{mn} \cd^2 R^{mn}
\\
& \qquad
		+ \frac{1}{1\,296} R^3
		+ \frac{1}{1\,080} R R_{mpqr} R_{mpqr}
		- \frac{1}{1\,080} R R_{mn} R_{mn}
		+ \frac{1}{336} R   \cd^2 R 
		\,.
\end{aligned}
\end{equation}

For second-order differential operators of the form \eqref{kad} we have, following
 \cite{Gilkey:1975iq,Avramidi:1990je,Avramidi:2000bm},
\begin{align} \label{app2}
\begin{aligned}
	b_6(\Delta_2)
=& \tr\Big[  	 
		- \frac{1}{60} \left( \cd_m \WW_{mn}  \right)^2
		+ \frac{1}{90} \WW_{mn} \WW_{nr} \WW_{rm}
		- \frac{1}{12} X \WW_{mn} \WW_{mn}
		+ \frac{1}{12} X \cd^2 X
		- \frac{1}{6} X^3  
\\
& \qquad
		+\frac1 {12} R X^2 
		- \frac{1}{72} R^2 X 
		 - \frac1{30} X \cd^2 R
		+ \frac{1}{180} X R_{mn}R_{mn}  
		- \frac{1}{180} X R_{mnrs}R_{mnrs} 
\\
	& \qquad
					+\frac{1}{72} R \WW _{mn} \WW^{mn} 
- \frac{1}{90} R_{mn} \WW^{m r}\WW^{rn} 
+ \frac{1}{180} R_{mnpq} \WW_{mn} \WW_{pq}
	+  \mathfrak E \cdot  \mathbbm 1 
		\Big] ,
\end{aligned}
\end{align}
where \(\1\) is the identity in the internal space, where \(\tr\) acts.

For fourth-order differential operators of the form \eqref{kad} we have (with \(V=V_{mm}\))
\begin{equation}\label{avdd2}
\begin{aligned}
b_6^{d=6}(\Delta_4) = & \tr \Big[ 
		 - \frac{1}{30}  \left( \cd_m \WW_{mn}  \right)^2
		+ \frac{1}{45} \WW_{mn} \WW_{np} \WW_{pm}
		+ \frac{1}{360} V_{mn} V_{np} V_{pm}
		+ \frac{1}{480} V_{mn} V_{mn} V
\\
& \qquad
		+\frac{1}{2\,880}V V V
		+ \frac{1}{30}V_{mn} \cd_{(n} \cd_{p)} V_{pm} 
		+\frac{1}{120} V_{mn} \cd^2 V_{mn}
		 - \frac{1}{40} V_{mn} \cd_{m} \cd_{n} V
\\
& \qquad
		+ \frac{1}{240} V \cd^2 V
		- \frac{1}{12}  V_{mn} V_{np}  \WW_{mp} 
		+  \frac{1}{6} \WW_{mn} \cd_{(m} \cd_{p)} V_{pn}
		+  \frac{1}{24} V \WW_{mn} \WW_{mn}
\\
& \qquad
		- \frac{1}{6}  V_{mn} \WW_{mp} \WW_{np}  
		- \frac{1}{3} \WW_{mn} \cd_{m} N_n 
		- \frac{1}{6}  V_{mn} \cd_{m} N_n 
		+ \frac{1}{12} V \cd_m N_m
		- \frac{1}{6}  N_m N_m
\\
& \qquad
		 - \frac{1}{12}   U V 
 			+\frac 1 {36} R \WW_{mn} \WW_{mn} 
			+ \frac 1 {144}  R V_{mn} V_{mn}
			+ \frac 1 {288}  R V V
			-\frac 1 6 RU
			+ \frac 1 {60}  R \cd^2 V 
\\
&  \qquad
			- \frac 1 {20}  R \cd_m \cd_n V_{mn}
			+ \frac 1 6 R \cd_m N_m
			- \frac 1 {36}  R_{mn} V_{np}V_{pm}
			- \frac 1 {72} R_{mn} V_{mn}V 
\\
& \qquad
			- \frac 1 {60}  R_{mn} \cd^2 V_{mn}
			- \frac 1 {12}  R_{mn} V_{np} \WW_{mp}
			+ \frac 1 {45} R_{mn} \WW_{mp} \WW_{np}
			+ \frac 1 {36}  R_{mnpq} V_{mp} V_{nq} 
\\
&  \qquad
			+ \frac 1 {90}  R_{mnpq} \WW_{mn} \WW_{pq} 
			+ \frac 1 {360}   R_{mnpq} R_{mnpq}  V   
			- \frac 1 {90}    R_{mpqk} R_{npqk}  V_{mn}
\\
&  \qquad
			- \frac 1{ 360}    R_{mn} R_{mn}  V 
			+ \frac 1 {144}    R^2  V 
			+ \frac 1 {45}     R_{mn} R_{mp}  V_{pn}  
			- \frac 1 {36}      R R_{mn}  V_{mn} 
\\
&  \qquad
			-\frac 1 {90}     R_{mnpq} R_{mp}  V_{nq} 
			 + 2 \mathfrak E  \cdot  \mathbbm 1 
					 \Big]\,.
\end{aligned}
\end{equation}  
The formula \eqref{avdd2} extends the result of \cite{Casarin:2019aqw} with the  present paper.

\subsection{A note on self-adjointness}\label{sadj}
Often we are interested in the operators of the form \eqref{kad} arising after path integration.
This operation typically projects on self-adjoint part, which imposes restrictions on the coefficient functions.
These conditions are \emph{not} automatically taken into account in expression for \(b_p\) such as \eqref{app2} and \eqref{avdd2}, which apply to generic operators, and have to be imposed by hand when isolating the differential operator in the quadratic part of the action.

For the  second order operator \(\Delta_2\) these translate on the requirement that \(X = X^\dagger\), where \(\dagger\) is the appropriate conjugation on the internal index structure (i.e.\ it is transposition or hermitian conjugation for real or complex fields respectively).

The discussion for \( \Delta_4 \) is slightly more subtle. It is convenient to rewrite the operator in the symmetric form
\begin{align}\label{zadd}
	\Delta_{4}
& =
	 \cd^4
			+ \cd_r  \hat{ V }_{rk}   \cd_k
			+ \hat{N}_k \cd_k
			+ \cd_k  \hat{N}_k 
			+ \hat{U},
& & 
	\hat{ V }_{rk}  = \hat{V}_{kr} \  , 
\end{align}
where the derivatives act on everything to their right. The relation with the non-symmetric form in \eqref{kad}  is  given by
\begin{equation}\label{zag}
{ V }_{mn} = \hat {V}_{mn}\ ,\quad\qquad 
{N}_m  =  \hat{N}_m +\frac12  \cd_m \hat{V}_{mn}\ ,
\qquad  \quad 
U  = \hat{U} + \cd_m \hat{N}_m \ . 
\end{equation}  
The form \eqref{zadd} is convenient because self-adjointness amounts to the conditions
\begin{equation}\label{zaf}
\hat {V}_{mn}= \hat {V}_{mn}^\dagger \ ,
\quad\qquad 
\hat {N}_{m} = -  \hat{N}_m ^\dagger \,,
\qquad  \quad 
\hat U  = \hat {U}^\dagger \,,
\end{equation} 
where again \(\dagger\) is the appropriate conjugation of the internal indices.

\section{Basis of the invariants for the decomposition \texorpdfstring{\(\Delta_4 = \Delta_+ \Delta_-\)}{D4=D+D-}}
\label{app::basis2}
In this appendix we list the basis of the invariants  used to study the decomposition \eqref{aaz}. We consider terms of \(\mathcal O (B^n)\), with \(n=1,2,4,5,6\). In total we find
45 elements; any other combination can be expressed in terms of these by integration by parts, use of Bianchi identities and other symmetry properties.
\begin{itemize}
\item \(\mathcal O (B^1)\):   19 elements
\begin{equation*}
\begin{gathered}
\cd_n \WW_{mn} R_{ma} B_a \,,\quad 
\WW_{mn} \cd_n  R_{ma} B_a   \,,\quad 
\cd_a  \WW_{mn}  R_{ma} B_n   \,,\quad 
\cd_m  \WW_{mn}  R  B_n  \,,
\\
\WW_{mn}  \cd_m R  B_n     \,,\quad 
\WW_{mn} \cd_a \WW_{mn}  B_a \,,\quad 
\WW_{mn} \cd_a  \WW_{ma}  B_n   \,,
\\ 
R_{mnac} \cd_c \WW_{mn} B_a \,,\quad 
R_{mnac}\WW_{ma}  \cd_n  B_c \,,\quad  
R_{mnac}\WW_{ma}  \cd_c  B_n\,,\quad  
R_{mnac}  \cd_r    R_{mnac}  B_r \,,
\\  
R_{mn}  \cd_r    R_{mn}  B_r \,,\quad 
R_{mr}  \cd_r    R_{mn}  B_n \,,\quad 
R_{mn}  \cd_n  R   B_m \,,\quad 
R  \cd_m    R  B_m  \,,
\\
R_{mnac}  \cd_c    R_{ma}  B_n \,,\quad 
R_{mnac}  \cd_n    R_{ma}  B_c \,,\quad  
\cd^2 \cd_m \WW_{mn} B_n \,,\quad 
\cd^2 R \cd_m B_m   \,.
\end{gathered} 
\end{equation*}

\item \(\mathcal O (B^2)\):  26 elements
\begin{equation*}
\begin{gathered}
\WW_{mn} \WW_{mn} B_a B_a     \,,\quad 
\WW_{ma} \WW_{mc} B_a B_c     \,,\quad 
R_{ma} \WW_{mc} B_a B_c     \,,\quad 
R_{mnac} \WW_{ma} B_n B_c     \,,
\\
R^2 B_a B_a     \,,\quad 
RR_{mn} B_m B_n     \,,\quad 
R_{ma}R_{mc}B_a B_c     \,,\quad 
R_{mnac} R_{mnac} B_r B_r     \,,\quad 
R_{mnac} R_{rnac} B_m B_r    \,, 
\\
\WW_{mn} B_m \cd^2 B_n   \,,\quad 
\WW_{ma} B_m \cd_a \cd_n B_n  \,,\quad  
\WW_{ma}\cd_a  B_m \cd_n B_n  \,,\quad  
\WW_{ma} B_n \cd_n \cd_a B_m  \,,
\\
R_{mnac} B_m B_a R_{nc}  \,,\quad  
B_a \cd^2 \cd^2 B_a \,,\quad
B_a \cd_a \cd^2 \cd_m B_m \,\quad
R_{mn} B_m \cd^2 B_n  \,,
\\
R_{mn} \cd_a B_m \cd_a B_n  \,,\quad  
R_{mn} B_a \cd_m\cd_n B_a  \,,\quad  
R_{mn} B_m \cd_n \cd_a B_a  \,,\quad  
R_{mn} B_a \cd_a \cd_n B_m\,,
\\
R B_m \cd_m \cd_n B_n  \,,\quad  
R \cd_m B_m \cd_n B_n  \,,\quad  
R B_a \cd^2 B_a  \,,\quad  
R \cd_a B_c \cd_a B_c  \,,\quad  
R_{mnac}  B_m \cd_n \cd_a B_c\,.
\end{gathered} 
\end{equation*}

\item \(\mathcal O (B^4)\): 8 elements
\begin{equation*}
\begin{gathered}
B_a B_a B_r \cd^2 B_r    \,,\quad 
B_a B_a \cd_m B_r \cd_m B_r    \,,\quad 
B_a \cd_m \cd_n B_a B_m B_n    \,,\quad 
B_a B_a B_m  \cd_m \cd_n B_n    \,,
\\
B_a B_a   \cd_m B_m  \cd_n B_n   \,,\quad 
B_a B_a   \cd_n B_m  \cd_m  B_n    \,,\quad 
R_{mn}  B_m B_n B_a  B_a \,,\quad 
R_{ }  B_m B_m B_a  B_a  \,.
\end{gathered} 
\end{equation*}

\item \(\mathcal O (B^5)\):    1 element
\begin{equation*} 
B_m B_m  B_a B_n \cd_n B_a 
\end{equation*}

\item \(\mathcal O (B^6)\): 1 element
\begin{equation*} 
(B_m B_m)^3
\end{equation*}

\end{itemize}

\section{Diagrammatic checks of \texorpdfstring{\(b^\mathrm{m}_6(\Delta_4)\)}{b6m(d4)}}
\label{app::diagr}

In this appendix we  compute  diagrammatically   some of the terms in \eqref{avd}-\eqref{ccs} as an independent consistency check.
We consider a free scalar in dimensional regularisation (\(d=6-2\varepsilon \)) with
\begin{equation}\label{yaaa}
S = \frac 12 \int\!  \sqrt g \ \phi \, \Delta_{4\phi}  \, \phi  
\,, 
\qquad\quad
\Gamma_\infty  =  \frac 1 {( 4 \pi)^3 \varepsilon} \int \! \sqrt{g} \ b_6( \Delta_{4\phi}  ) \,,
\end{equation}
where \(\Delta_{4\phi}\) has the structure \eqref{iad} with spacetime connection  only and with  \(V,N,U\) being spacetime  covariant functions.

To set the perturbative expansion in powers of the external fields we use   \( h_{mn} = g_{mn } - \delta_{mn}  \) and for simplicity we assume  \(h_{mm}= h_{mn} \delta_{mn}=0\). 
We construct the diagrams for the following three correlators:
  \( \braket{ U h } \)  (giving \(c_4\)),
  \(\braket{Nh }\)  (giving \(c_7\)), 
  \( \braket{Vh}\) (giving \(c_{5}\), \(c_{6}\), \(c_{10}\)).
The results of this calculation are all in  agreement with the solution in \eqref{ccs}.

To diagrammatically compute the divergent part of the effective action from \eqref{yaaa} we need the free propagator
\begin{equation}\label{yaa}
\braket{\phi(p) \, \phi (-p)   } =    \frac{1}{p^4} 
\end{equation}
and the following vertices, 
\begin{equation}\label{yab}
\begin{aligned}
S_h &=  \frac 12 \int\!   h_{mn}(-p-q) H_{mn}(p,q) \phi(q) \phi(q)  \,,\qquad
&
H_{mn}(p,q) &= p_{(m} q_{n)} (p^2 + q^2)   \,;
\\
S_V& =  \frac 12 \int\!   V_{mn}(p-q)   W_{mn}(p,q) \phi(q) \phi(q)    \, , 
&
W_{mn}(p,q) & =  - \frac12 ( q_mq_n + p_mp_n)  \,;
\\ 
S_N& =  \frac 12 \int\!   N_{m}(p-q)  M_m(p,q) \phi(q) \phi(q)    \,,
&
M_m(p,q)& =i (p_m+q_m)   \,;
\\
S_U &=  \frac 12 \int\!   U(-p-q) \phi(p) \phi(q)  \,.
\end{aligned}
\end{equation}

To the terms under considerations  only a single two-propagator diagram contributes, and the  corresponding terms in the effective action are found to be 
\begin{equation}\label{yba}
\begin{aligned}
\braket{V_{rs} (q) \  h_{mn}(-q)  }&=
-\frac12\int \!  \frac{d^dp}{(2\pi)^d} \frac{1}{p^4 (q-p)^4  } H_{mn}(p,q-p) W_{rs}(p-q,-p) \,,
\\
\braket{  N_r (q)\    h_{mn}(-q) }&=
-\frac12\int  \! \frac{d^dp}{(2\pi)^d} \frac{1}{p^4 (q-p)^4  } H_{mn}(p,q-p) M_{r}(p-q,-p) \,,
\\
\braket{  U (q)   \ h_{mn}(-q) } & =
-\frac12\int  \! \frac{d^dp}{(2\pi)^d} \frac{1}{p^4 (q-p)^4  } H_{mn}(p,q-p) \,.
\end{aligned}
\end{equation}
The loop integrals  can be evaluated using standard two-propagator technology (see e.g.~\cite{Casarin:2021fgd}) and the divergent parts read
\begin{align}\nonumber
\braket{V_{rs} (q) \  h_{mn}(-q)  }_\infty &=
\frac{1}{240 (4\pi)^3 \varepsilon} 
\left[
6 q_{r} q_{s} q_{m} q_{n} -  \delta_{rm} \delta_{sn} q^4 + 2 \delta_{rn} q_{s} q_{m} q^2 - 2 \delta_{rs} q_{m} q_{n} \
q^2
\right],
\\
\braket{  N_r (q)\    h_{mn}(-q) }_\infty &=
\frac{ i }{12 (4\pi)^3 \
\varepsilon} 
q_r  q_{m} q_{n}  \,,
\\\nonumber
\braket{  U (q)   \ h_{mn}(-q) }_\infty  & =
- \frac{1}{12 (4\pi)^3 \varepsilon} 
 q_{m} q_{n}  \,,
\end{align}
which correctly reproduce the values for \(c_4,c_5,c_6,c_7,c_{10}\).

\bibliography{biblio} 

\bibliographystyle{JHEP}
\end{document}